\documentclass[twocolumn,runningheads]{svjour2}
\smartqed  % flush right qed marks, e.g. at end of proof
\usepackage{graphicx}
\usepackage{mathptmx}      % use Times fonts if available on your TeX system
\journalname{Astrophysics and Space Science}

\usepackage{graphicx}
\usepackage{floatflt, graphicx}
\usepackage{amssymb}
\usepackage{subfigure}

\begin{document}

\title {\bf Cosmic Neutrinos from the Sources of Galactic and Extragalactic Cosmic Rays}

\author{Francis Halzen}

\authorrunning{F. Halzen}

\institute{F. Halzen \at
           Department of Physics,
           University of Wisconsin,
           Madison, WI 53706
           Tel.: 608 262 2667\\
           Fax: 608 262 8628\\
           \email{halzen@icecube.wisc.edu}
}

\date{Received: date / Accepted: date}

\maketitle

\begin{abstract}
Although kilometer-scale neutrino detectors such as IceCube are discovery instruments, their conceptual design is very much anchored to the observational fact that Nature produces protons and photons with energies in excess of $10^{20}$\,eV and $10^{13}$\,eV, respectively. The puzzle of where and how Nature accelerates the highest energy cosmic particles is unresolved almost a century after their discovery. From energetics considerations we anticipate order $10\sim100$ neutrino events per kilometer squared per year pointing back at the source(s) of both galactic and extragalactic cosmic rays. In this context, we discuss the results of the AMANDA and IceCube neutrino telescopes which will deliver a kilometer-square-year of data over the next 3 years.

%\PACS{First \and Second \and More}

\end{abstract}

\section{Introduction}

Ambitious projects have been launched to extend conventional astronomy beyond wavelengths of $10^{-14}$\,cm, or GeV photon energy. Besides gamma rays, protons (nuclei), neutrinos and gravitational waves will be explored as astronomical messengers probing the extreme Universe. The challenges are considerable:
\begin{itemize}
\item Protons are relatively abundant, but their arrival directions have been scrambled by magnetic fields.
\item $\gamma$-rays do point back to their sources, but are absorbed at TeV-energy and above on cosmic background radiation.
\item neutrinos propagate unabsorbed and without deflection throughout the Universe but are difficult to detect.
\end{itemize}
Therefore, multi-messenger astronomy may not just be an advantage, it may be a necessity for solving some of the outstanding problems of astronomy at the highest energies such as the identification of the sources of the cosmic rays, the mechanism(s) triggering gamma ray bursts and the particle nature of the dark matter.

We here discuss the detection of neutrinos associated with the observed fluxes of high energy cosmic rays and gamma rays. We will show that the anticipated fluxes point at the necessity of commissioning kilometer-scale neutrino detectors. Though ambitious, the scientific case is compelling because neutrinos will reveal the location of the source(s) and represent the ideal tool to study the black holes powering the cosmic accelerator(s).

Soon after the discovery in the mid-fifties that neutrinos were real particles and not just mathematical constructs of theorists' imagination, the idea emerged that they represent ideal cosmic messengers\cite{reines}. Because of their weak interactions, neutrinos reach us unimpeded from the edge of the Universe and from the inner reaches of black holes. The neutrino telescopes now under construction have the capability to detect neutrinos with energies from a threshold of $\sim 10$\,GeV to, possibly, $ \sim 10^2$\,EeV, the highest energies observed. Their telescope range spans more than 10 orders of magnitude in wavelengths smaller than $10^{-14}$\,cm. This is a reach equivalent to that of a hypothetical astronomical telescope sensitive to wavelengths from radio to X-rays. Above $10^5$\,TeV the observations are free of muon and neutrino backgrounds produced in cosmic ray interactions with the Earth's atmosphere. Each neutrino is a discovery.\footnote{We will use  GeV$=10^9$\,eV, TeV$=10^{12}$\,eV,  PeV$=10^{15}$\,eV and EeV$=10^{18}$\,eV units of energy}

The real challenge of neutrino astronomy is that kilometer-scale neutrino detectors are required to do the science. The first hint of the scale of neutrino telescopes emerged in the nineteen seventies from theoretical studies of the flux of neutrinos produced in the interactions of cosmic rays with microwave photons, the so-called Greissen-Zatsepin-Kuzmin or GZK neutrinos. Since then the case for kilometer-size instruments has been strengthened\cite{PR} and the possibility of commissioning such instruments demonstrated\cite{ice3}. In fact, if the neutrino sky were within reach of smaller instruments, it would by now have been revealed by the first-generation AMANDA telescope. It has been taking data since 2000 with a detector of 0.01--0.08\,km$^2$ telescope area, depending on the sources\cite{pune}.

Given the size of the detector required, all efforts have concentrated on transforming large volumes of natural water or  ice into Cherenkov detectors. They reveal the secondary muons and electromagnetic and hadronic showers initiated in neutrino interactions inside or near the detector. Because of the long range of the muon, from kilometers in the TeV range to tens of kilometers at the highest energies, neutrino interactions can be identified far outside the instrumented volume. Adding to the technological challenge is the requirement that the detector be shielded from the abundant flux of cosmic ray muons by deployment at a depth of typically several kilometers. After the cancellation of a pioneering attempt\cite{water} to build a neutrino telescope off the coast of Hawaii, successful operation of a smaller instrument in Lake Baikal\cite{baikal} bodes well for several efforts to commission neutrino telescopes in the Mediterranean\cite{water,emigneco}. We will here mostly concentrate on the construction and first four years of operation of the AMANDA telescope\cite{pune,nature} which has transformed a large volume of natural deep Antarctic ice into a Cherenkov detector. It represents a first-generation telescope as envisaged by the DUMAND collaboration over 20 years ago and a proof of concept for the kilometer-scale IceCube detector, now under construction.

Even though neutrino ``telescopes" are designed as discovery instruments covering a large dynamic range, be it for particle physics or astrophysics, their conceptual design is very much anchored to the observational fact that Nature produces protons and photons with energies in excess of $10^{20}$\.eV and $10^{13}$\,eV, respectively.  In this paper we will review how cosmic ray and TeV gamma ray observations set the scale of cosmic neutrino fluxes.

\section{Cosmic Neutrinos Associated with Extragalactic Cosmic Rays}
 
Cosmic accelerators produce particles with energies in excess of $10^8$\,TeV; we do not know where or how. The flux of cosmic rays observed at Earth is sketched in Fig.\,1a,b\cite{gaisseramsterdam}. The  energy spectrum follows a broken power law. The two power laws are separated by a feature dubbed the ``knee"; see Fig.\,1a. Circumstantial evidence exists that cosmic rays, up to perhaps EeV energy, originate in galactic supernova remnants. Any association with our Galaxy disappears in the vicinity of a second feature in the spectrum referred to as the ``ankle". Above the ankle, the gyroradius of a proton in the galactic magnetic field exceeds the size of the Galaxy and it is generally assumed that we are  witnessing the onset of an extragalactic component in the spectrum that extends to energies beyond 100\,EeV. Experiments indicate that the highest energy cosmic rays are predominantly protons or, possibly, nuclei. Above a threshold of 50 EeV these protons interact with cosmic microwave photons and lose energy to pions before reaching our detectors. This is the GZK cutoff that limits the sources to our local supercluster. 

%% fig.1
\begin{figure*}[]
\centering\leavevmode
\includegraphics[width=0.8\textwidth]{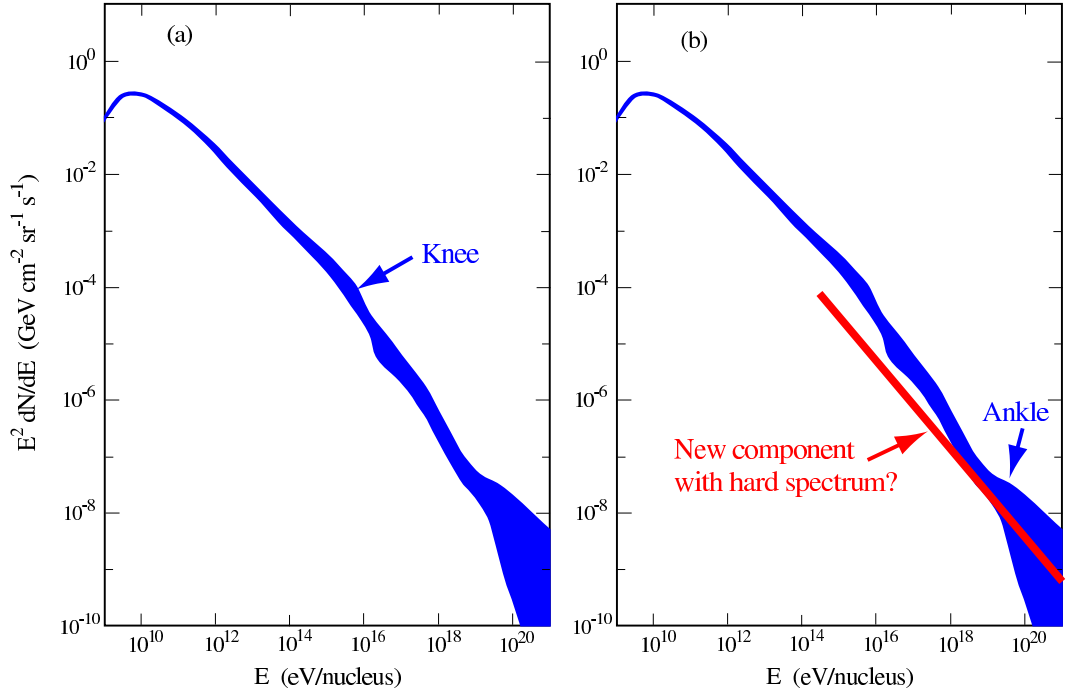}
\caption{At the energies of interest here, the cosmic ray spectrum consists of a sequence of 3 power laws. The first two are separated by the ``knee" (left panel), the second and third by the "ankle". There is evidence that the cosmic rays beyond the ankle are a new population of particles produced in extragalactic sources; see right panel.}
\label{knee-ankle}
\end{figure*}

Models for the origin of the highest energy cosmic rays fall into two categories, top-down and bottom-up. In top-down models it is assumed that the cosmic rays are the decay products of cosmological remnants or topological defects associated, for instance, with Grand Unified theories with unification energy $M_{GUT} \sim 10^{24}\rm\,eV$. These models predict neutrino fluxes most likely within reach of first-generation telescopes such as AMANDA, and certainly detectable by future kilometer-scale neutrino observatories\cite{semikoz}. They have not been observed.

In bottom-up scenarios it is assumed that cosmic rays originate in cosmic accelerators. Accelerating particles to TeV energy and above requires massive bulk flows of relativistic charged particles. These are likely to originate from the exceptional gravitational forces  in the vicinity of black holes. Gravity powers large electric currents that create the opportunity for particle acceleration by shocks, a mechanism familiar from solar flares where particles are accelerated to $10$\,GeV. It is a fact that black holes accelerate electrons to high energy; astronomers observe them indirectly by their synchrotron radiation. We know that they can accelerate protons because we detect them as cosmic rays. Being charged, the protons are deflected by interstellar magnetic fields and therefore do not reveal their sources. Hence the cosmic ray puzzle.

Examples of candidate black holes include the dense cores of exploding stars, inflows onto supermassive black holes at the centers of active galaxies and annihilating black holes or neutron stars. Before leaving the source, accelerated particles pass through intense radiation fields or dense clouds of gas surrounding the black hole. This results in interactions producing pions decaying into secondary photons and neutrinos that accompany the primary cosmic ray beam as illustrated in Fig.\,2.
%
%% fig.2
\begin{figure*}[]
\centering\leavevmode
\includegraphics[width=0.6\textwidth]{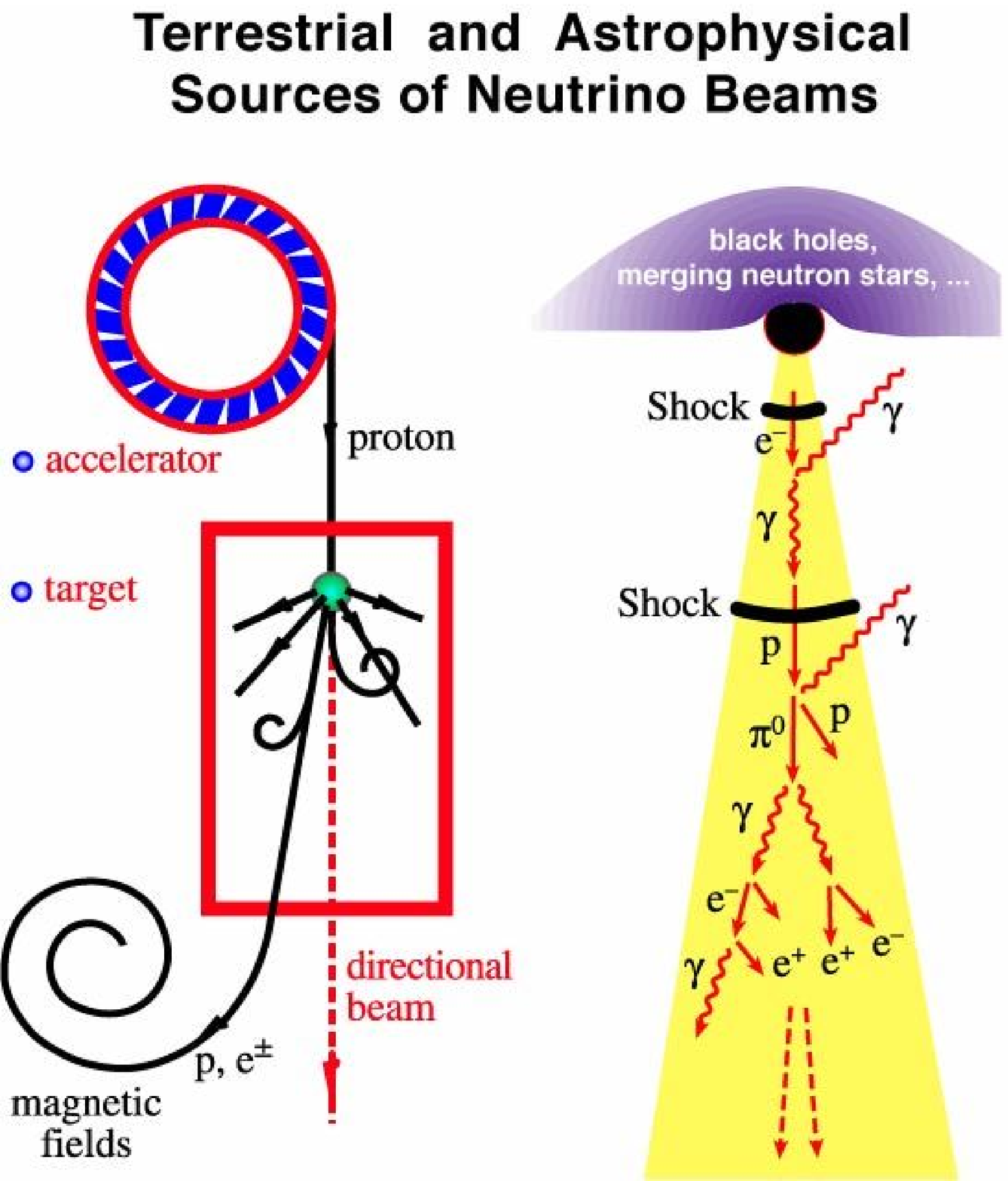}
\caption{Cosmic beam dump exits: sketch of cosmic ray accelerator producing photons. The charged pions that are inevitably produced along with the neutral pions will decay into neutrinos.}
\label{nubeams}
\end{figure*}
 How many neutrinos are produced in association with the cosmic ray beam? 
 The answer to this question provides one rationale for building
 kilometer-scale  neutrino detectors~\cite{PR}. For orientation,
 consider a neutrino beam produced at an accelerator laboratory. 
 Here the target and the beam dump absorb all parent protons as
 well as the secondary electromagnetic and hadronic showers. 
 Only neutrinos exit the dump. If Nature constructed such a
 ``hidden source'' in the heavens, conventional astronomy would
 not reveal it.  Cosmic ray sources must be at least partially
 transparent to protons.  Sources transparent only to neutrinos
 may exist, but they cannot be cosmic-ray sources. 

 A generic ``transparent'' source can be imagined as follows: protons
 are accelerated in a region of high magnetic fields where they
 interact with photons and generate neutral and charged pions.
 The most important process is 
 $p + \gamma \rightarrow \Delta^+ \rightarrow \pi^0 + p$
 and
 $p + \gamma \rightarrow \Delta^+ \rightarrow \pi^+ + n$.
 While the secondary protons may remain trapped in the acceleration
 region, roughly equal numbers of neutrons and decay products of 
 neutral and charged pions escape. The energy escaping the source
 is therefore distributed among cosmic rays, gamma rays and
 neutrinos produced by the decay of neutrons, neutral pions and
 charged pions, respectively. 
 The neutrino flux from a generic transparent cosmic ray source is
 often referred to as the Waxman-Bahcall flux~\cite{wb1}. 
 It is easy to calculate and the derivation is revealing.

 Figure 1b shows a fit to the observed spectrum above the
 ``ankle" that can be used to derive the total energy in extragalactic
 cosmic rays. The flux above the ankle is often summarized as
 ``one $10^{19}$\,eV particle per kilometer square per year per
 steradian". This can be translated into an energy flux
\begin{eqnarray*}
E \left\{ E{dN\over dE} \right\}&=& {10^{19}\,{\rm eV} \over \rm (10^{10}\,cm^2)(3\times 10^7\,sec) \, sr}\\[0.3 cm]
&=& 3\times 10^{-8}\rm\, GeV\ cm^{-2} \, s^{-1} \, sr^{-1} \,.
\end{eqnarray*}
From this we can derive the energy density $\rho_E$ in cosmic rays using the relation that flux${}={}$velocity${}\times{}$density, or
\[
4\pi \int  dE \left\{ E{dN\over dE} \right\} =  c\rho_E\,.
\]
We obtain
\[
\rho_E = {4\pi\over c} \int_{E_{\rm min}}^{E_{\rm max}} {3\times 10^{-8}\over E} dE \, {\rm {GeV\over cm^3}} \simeq 10^{-19} \, {\rm {TeV\over cm^3}} \,,
\]
taking the extreme energies of the accelerator(s) to be\\$E_{\rm max} / E_{\rm min} \simeq 10^3$.

The energy content derived ``professionally" by integrating the spectrum in Fig.~2b assuming an $E^{-2}$ 
energy spectrum, typical of shock acceleration, with a GZK cutoff  is $\sim 3 \times 10^{-19}\rm\,erg\ cm^{-3}$. 
This is within a factor of our back-of-the-envelope estimate (1\,TeV = 1.6\,erg). 
The power required for a population of sources to generate this energy density over 
the Hubble time of $10^{10}$\,years is $\sim 3 \times 10^{37}\rm\,erg\ s^{-1}$ per (Mpc)$^3$ or, 
as often quoted in the literature, $\sim 5\times10^{44}\rm\,TeV$ per (Mpc)$^3$ per year. 
This works out to\cite{TKG}
\begin{itemize}
\item $\sim 3 \times 10^{39}\rm\,erg\ s^{-1}$ per galaxy,
\item $\sim 3 \times 10^{42}\rm\,erg\ s^{-1}$ per cluster of galaxies,
\item $\sim 2 \times 10^{44}\rm\,erg\ s^{-1}$ per active galaxy, or
\item $\sim 2 \times 10^{52}$\,erg per cosmological gamma ray burst.
\end{itemize}
The coincidence between these numbers and the observed output in electromagnetic energy of these 
sources explains why they have emerged as the leading candidates for the cosmic ray accelerators. 
The coincidence is consistent with the relationship between cosmic rays and photons built into the ``transparent" 
source. In the photoproduction processes roughly equal energy goes into the secondary neutrons, neutral 
and charged pions whose energy ends up in cosmic rays, gamma rays and neutrinos, respectively.

We therefore conclude that the same energy density  of $\rho_E \sim 3 \times 10^{-19}\rm\,erg\
 cm^{-3}$, observed in cosmic rays and electromagnetic energy, ends up in neutrinos with a spectrum $E_\nu dN / dE_{\nu}  \sim E^{-\gamma}\rm\, cm^{-2}\, s^{-1}\, sr^{-1}$ that continues up to a maximum energy $E_{\rm max}$. The neutrino flux follows from the relation
%
%\begin{eqnarray}
$ \int E_\nu dN / dE_{\nu}  =  c \rho_E / 4\pi  $.
%\end{eqnarray}
%
For $\gamma = 1$ and $E_{\rm max} = 10^8$\,GeV, the generic source of the highest energy cosmic rays produces a flux of $ {E_\nu}^2 dN / dE_{\nu}  \sim 5 \times 10^{-8}\rm\, GeV \,cm^{-2}\, s^{-1}\, sr^{-1} $.

There are several ways to sharpen  this qualitative prediction:
\begin{itemize} 
\item The derivation fails to take into account that there are more UHE
  cosmic rays in the Universe than observed at Earth because
  of the GZK-effect and it also neglects the evolution of the
  sources with redshift. This increases the neutrino flux, which
  we normalized to the observed spectrum only, by a factor
  $d_H/d_{\rm CMB}$, the ratio of the Hubble radius to the
  average attenuation length of the cosmic rays propagating in
  the cosmic microwave background.
\item For proton-$\gamma$ interactions muon neutrinos (and
  antineutrinos) receive only 1/2 of the energy of the charged pion in
  the decay chain $\pi^+\rightarrow \mu^+ +\nu_{\mu}\rightarrow e^+
  +\nu_e +\bar{\nu}_{\mu} +\nu_{\mu}$ assuming that the energy is
  equally shared between the 4 leptons. Furthermore half the muon
  neutrinos oscillate into tau neutrinos over cosmic distances. In further calculations we will focus on the muon flux here.
\end{itemize}
In summary,
\begin{equation}
E_\nu{dN_\nu\over dE_\nu} = {1\over2} \times {1\over2} 
\times E{dN_{\rm CR}\over dE} \times {d_H\over d_{\rm CMB}} 
\simeq E{dN_{\rm CR}\over dE}
\end{equation}
 In practice, the corrections approximately cancel. The precise value of the energy where the transition from galactic to extragalactic sources occurs represents another source of uncertainty that has been extensively debated~\cite{ringwald}. A transition at a lower energy significantly increases the energy in the extragalactic component and results in an enhancement of the associated neutrino flux.

Waxman and Bahcall referred to their flux as a bound in part because in
reality more energy is transferred to the neutron than to the charged
pion in the source, in the case of the photoproduction reaction $p +
\gamma \rightarrow \Delta^+ \rightarrow \pi^+ + n$ four times more.
Therefore
\begin{equation}
E_\nu{dN_\nu\over dE_\nu} = {1\over4} E{dN_{\rm CR}\over dE} \, .
\end{equation}
 In the end we estimate that the muon-neutrino flux associated with
 the sources of the highest energy cosmic rays is loosely confined
 to the range
 \begin{equation}{E_\nu}^2 dN / dE_{\nu}= 1\sim 5 \times 10^{-8}\rm\,
 GeV \,cm^{-2}\, s^{-1}\, sr^{-1}
\end{equation} 
 depending on the cosmological
 evolution of the cosmic ray sources. Model calculations assuming that active galaxies or gamma-ray bursts are the actual sources of cosmic rays yield event rates similar to the generic energetics estimate presented.

The anticipated neutrino flux thus obtained has to be compared with the limit of\\
$8.9 \times 10^{-8}\rm\, GeV\ cm^{-2}\, s^{-1}\,sr^{-1}$ reached after the first 
4 years of operation of the completed AMANDA detector in 2000--2003~\cite{hill}.  On the other hand, after three years of operation IceCube will 
reach a diffuse flux limit of
\begin{equation}
E_{\nu}^2 dN / dE_{\nu} = 2\,{\sim}\, 7 \times 10^{-9}\rm\,GeV \,cm^{-2}\, s^{-1}\, sr^{-1}.
\end{equation} 
The exact value of the IceCube sensitivity depends on the magnitude of the dominant high energy 
neutrino background from the prompt decay of atmospheric charmed particles\cite{ice3}. The level 
of this background is difficult to anticipate theoretically and little accelerator data is available 
in the energy and Feynman-x range of interest\cite{ingelman}.

%% fig.3
\begin{figure*}[]
\centering\leavevmode
\includegraphics[width=0.8\textwidth]{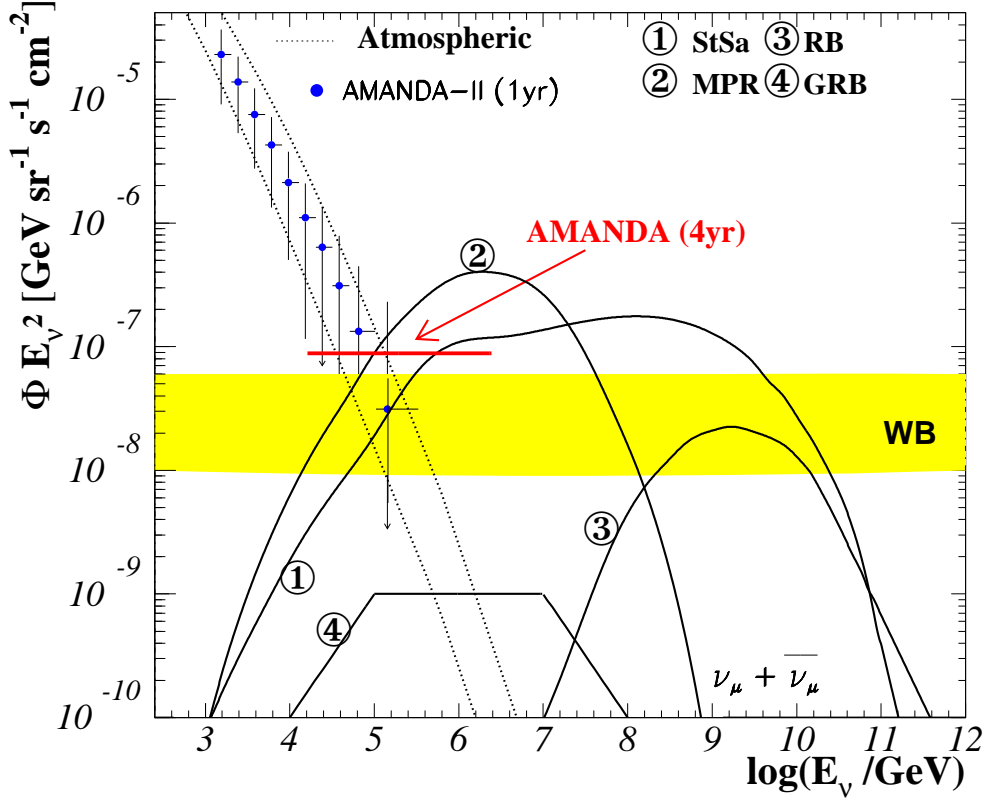}
\caption{Our estimate of the flux of neutrinos associated with the
 sources of the highest energy cosmic rays (the shaded range labeled WB)
 is compared to the limits established by the AMANDA experiment reached with
 800 days of data\cite{hill}. AMANDA's sensitivity is within a factor of 2 of the most optimistic predictions. Also shown are fluxes predicted by specific models
 of cosmic ray accelerators: active galaxies labeled StSa\protect\cite{agn}
 and MPR\protect\cite{MPR}, GRB\protect\cite{guetta} and
 the diffuse flux produced by cosmic ray producing active galaxies on
 microwave photons\protect\cite{RB} labelled RB. Data for the background
 atmospheric neutrino flux are from the AMANDA experiment. The IceCube experiment will be sensitive to all predictions after a few years of operation of the full detector. It has sensitivity to the larger fluxes by operating the partially completed detector that already now exceeds AMANDA in instrumented volume.}
\label{fig:diffuse_incl_osc_gs_ai.eps}
\end{figure*}

The observed event rate is obtained by folding the cosmic flux predicted with the probability that 
the neutrino is actually detected in a high energy neutrino telescope; only one in a million neutrinos 
of TeV energy interact and produce a muon that reaches the detector. This probability is given by the 
ratio of the muon and neutrino interaction lengths in the detector medium, 
$\lambda_\mu / \lambda_\nu$\cite{PR} and therefore depends on energy. For the flux range estimated above we anticipate 100--500 detected muon neutrinos per km$^2$ per year. Here the lower value represents the more realistic estimate. It will be further reduced if we assume a steeper spectrum. On the other hand, given that IceCube's effective area for muon neutrinos exceeds 1\,km$^2$ and that equal fluxes of electron and tau neutrinos are expected, a neutrino signal at the ``Waxman-Bahcall" level could result in the observation of several hundred high-energy neutrinos of extraterrestrial origin per year in IceCube~\cite{ice3}. 

Gamma ray bursts (GRB), outshining the entire Universe for the duration of the burst, are perhaps the best motivated sources of high-energy neutrinos\cite{waxmanbahcall,mostlum2,mostlum3}. The collapse of massive stars to a black hole has emerged as the likely origin of the ``long" GRB with durations of tens of seconds. In the collapse a fireball is produced which expands with a highly relativistic velocity powered by radiation pressure. The fireball eventually runs into the stellar material that is still accreting onto the black hole. If it successfully punctures through this stellar envelope the fireball emerges to produce a GRB. While the energy transferred to highly relativistic electrons is thus observed in the form of radiation, it is a matter of speculation how much energy is transferred to protons.

The assumption that GRB are the sources of the highest energy cosmic rays does determine the energy of the fireball baryons. Accommodating the observed cosmic ray spectrum of extragalactic cosmic rays requires roughly equal efficiency for conversion of fireball energy into the kinetic energy of protons and electrons. In this scenario the production of neutrinos of 100--1000\,TeV energy in the GRB fireball is a robust prediction because neutrinos are inevitably produced in interactions of accelerated protons with fireball photons. Estimates of the flux\cite{guetta} point again at the necessity of a kilometer-cubed neutrino detector, in agreement with the generic energetics estimates previously presented. Studies of active galaxies as sources of cosmic rays lead to similar conclusions\cite{agn}.

The case for kilometer-scale detectors also emerges from consideration of  ``guaranteed'' cosmic fluxes. Neutrino fluxes are guaranteed when both the accelerator and the pion producing target material can be identified. The extragalactic cosmic rays produce $ \sim$ 1 event per km$^2$\,year in interactions with cosmic microwave photons\cite{cos}. Galactic cosmic rays interact with hydrogen in the disk to generate an observable neutrino flux in a kilometer-scale detector\cite{plane}. Evidence has been accumulating that young supernova remnants are the sources of the galactic cosmic rays; conclusive evidence is still missing. Neutrino observations can be the answer as we will review in the next section.

\section{Cosmic Neutrinos Associated with Galactic Cosmic Rays}

 In the previous section we made an estimate of the neutrino flux
 from generic accelerators producing the highest energy cosmic rays.
 We can perform a similar analysis for the galactic cosmic rays by
 calculating the energy density corresponding to the flux shown in
 Fig.\,1a. The answer is that $\rho_E \sim 10^{-12}$\,erg\,cm$^{-3}$.
 This is also the value of the corresponding energy density $B^2/8\pi$
 of the microgauss magnetic field in the galaxy. The power needed to maintain
 this energy density is $10^{-26}$\,erg/cm$^3$s given that the
 average containment time of the cosmic rays in our galaxy is
 $3\times10^6$\,years. For a nominal volume of the galactic disk
 of $10^{67}$\,cm$^3$ this requires an accelerator delivering
 $10^{41}$\,erg/s. This happens to be 10\% of the power produced by supernovae releasing
 $10 ^{51}\,$erg every 30 years. The coincidence is the basis for
 the idea that shocks produced by supernovae expanding into the
 interstellar medium are the origin of the galactic cosmic
 rays.
 
Can we observe neutrinos pointing back at the accelerators of the galactic cosmic rays? The conversion of the $10 ^{50}\,$erg of energy into particle acceleration is believed to occur by diffusive shock acceleration in the young (1000--10,000 years) remnant expanding into the interstellar medium. If high energy cosmic rays are indeed associated with the remnant, they will interact with hydrogen atoms in the interstellar medium to produce pions that decay into roughly equal numbers of photons and neutrinos. These may provide us with indirect  evidence for cosmic ray acceleration. The observation of these pionic gamma rays has been one of the motivations for neutrino as well ground-based TeV-energy astronomy.

Whereas the details are complex and predictions can be treacherous, a simple estimate of the gamma ray flux associated with a supernova remnant can be made following Aharonian, Drury and Volk\cite{ADV94}. Within the precision of the astrophysics it is safe to assume that an identical flux of neutrinos is produced-- no need for sophistication here. The emissivity in pionic gamma rays produced by a density of protons $n_p$ interacting with a density of hydrogen atoms n is
\begin{eqnarray}
Q_\gamma (> 1\,TeV) &=& c <{E_\pi \over E_p}> \sigma_{pp}\, n\, n_p(>1\,TeV)\\
&=& c <{E_\pi \over E_p}> {\lambda_{pp}}^{-1}\, n_p(>1\,TeV),
\end{eqnarray}
or
\begin{equation}
Q_\gamma (> 1\,TeV) \simeq 10^{-29} \,{\rm photons\over \rm s\,cm^3}\, ({n \over \rm 1\,cm^{-3}}).
\end{equation}
The emissivity of photons is simply proportional to the density of cosmic rays $n_p(>1\,TeV)$ ($\simeq4\times10^{-14}\, cm^{-3}$ for energy in excess of 1 TeV) and the target density $n$ of hydrogen atoms. The proportionality factor is determined by particle physics: $<E_\pi/E_p> \sim 0.2$ is the average energy of the secondary pions relative to the cosmic ray protons and $\lambda_{pp}= (n\sigma_{pp})^{-1}$ is the proton interaction length ($\sigma_{pp} \simeq 40$\,mb) in a density n of hydrogen atoms. (We here assumed a generic $E^{-2}$ spectrum of the protons, for different spectral indices the quantity $<E_\pi/E_p>$ is generalized to the spectrum-weighted moments for pion production by nucleons\cite{gaisser}.)

The total luminosity in gamma rays is given by
\begin{equation}
L_{\gamma} (> 1\,TeV)  = Q_\gamma\, {W \over \rho} \simeq 10^{33}\, \rm photons\, s^{-1}.
\end{equation}
The density of protons from a supernova converting a total kinetic energy W of $10^{50}$\,erg to proton acceleration is approximately given by $W/\rho$, where we will assume that the density in the remnant is not very different from the ambient energy density $\rho \sim 10^{-12}$\,erg\,cm$^{-3}$ of galactic cosmic rays. This approximation is valid for young remnants in their Sedov phase.

We thus predict a  rate of TeV photons from a supernova at a distance d of 1\,kpc of
\begin{eqnarray*}
{dN_{events}\over \rm d(lnE)} (&>&\,E) = {L_\gamma \over 4\pi d^2}\\& \simeq& \,10^{-11} \: ({\rm photons\over \rm cm^2\,s}) ({W_{CR}\over \rm 10^{50}\,erg}) ({n\over \rm 1\,cm^{-3}}) ({d\over \rm 1\,kpc})^{-2}.
\label{galactic1}
\end{eqnarray*}
Each TeV gamma ray is accompanied by a neutrino from a charged pion and we therefore anticipate an event rate of 3 detected neutrinos per decade of energy per km$^2$ year, a result readily obtained from the relation
\begin{equation}
{dN_{events}\over \rm d(lnE)} (>\,E) = 10^{-11} \: ({\rm neutrinos\over \rm cm^2 \: s}) \: \rm area \: \rm time \: ({\lambda_\mu\over \rm \lambda_\nu}),
\label{galactic2}
\end{equation}
where the last factor represents, as before, the probability that the neutrino is detected. It is approximately $10^{-6}$ for the TeV energy considered here. From several such sources distributed over the galactic plane IceCube may detect a flux of neutrinos similar, possibly smaller,  than the one associated with extragalactic sources.

This estimate may be somewhat optimistic because we assumed that the sources extend to 100\,TeV with an $E^{-2}$ spectrum. If the spectrum cuts off around 10 TeV detection becomes more challenging because the flux reaches the level of the cosmic ray background. On the other hand, if the ``knee" at 1000\,TeV represents the end of the galactic cosmic ray spectrum, then some of the sources must produce 100\,TeV secondaries.

This prediction is credible because the number of TeV photons predicted coincides with observations of the supernova remnant RX J1713.7-3946 by the H.E.S.S. array of atmospheric Cherenkov telescopes\cite{hess}. H.E.S.S. may thus have identified the first site where protons are accelerated to energies typical of the main component of the galactic cosmic rays. Although the resolved image of the source (the first ever at TeV energies!) reveals TeV gamma ray emission from the whole supernova remnant, it shows a clear increase of the flux in the directions of known molecular clouds. This is suggestive of protons, shock accelerated in the supernova remnant, interacting with the dense clouds to produce neutral pions that are the source of the observed increase of the TeV photon signal. The image shows filaments of high magnetic fields consistent with the requirements for acceleration to the energies observed. Furthermore, the high
 statistics data for the flux are power-law behaved over a large range of energies without any indication of a cutoff characteristic of synchrotron or inverse-Compton sources. Follow-up observations of the source in radio-waves and X-rays have failed to identify the population of electrons required to generate TeV photons by purely electromagnetic processes; for a detailed discussion see \cite{hiraga}.
 
On the theoretical side, the large B-fields suppress the ratio of photons produced by the inverse Compton relative to the synchrotron. Fitting the data by purely electromagnetic processes is therefore challenging but, apparently, not impossible\cite{hiraga}. A similar extended source of TeV gamma rays tracing the density of molecular clouds has been identified near the galactic center. Protons apparently accelerated by the remnant HESS J1745-290 diffuse through nearby molecular clouds to produce a signal of TeV gamma rays that trace their density\cite{GC}. Detecting this source in neutrinos will be challenging because it is relatively weak (its TeV luminosity is only or order 0.1 Crab), because of its larger distance compared to RX J1713.7-3946 and because it is not a point but extended source\cite{kistler,kappes}. On the other hand, the sources discovered by Milagro in the Cygnus region are more luminous, their spectrum extends to higher energies and they are relatively nearby\cite{goodman}. We are looking forward to a detailed measurement of the Milagro spectrum which is likely to translate into a detectable neutrino flux in IceCube. The hotspot in the Milagro map of the galactic plane represents a flux of 1 Crab above 12.5\,TeV\cite{goodman}. By itself it yields $4\sim17$\, neutrino events above a detector threshold of $0.1\sim1$\,TeV. We here assumed a spectrum of $E^{-2.6}$, but the result is relatively insensitive to this assumption. The diffuse flux surrounding the Milagro source represents an additional flux of 3 Crab spread over 0.02 steradian. IceCube data should have no problem exposing the smoking gun for the sources of galactic cosmic rays.

So far H.E.S.S. has not claimed the discovery of pionic gamma rays and finding neutrinos as a smoking gun for cosmic ray acceleration in supernova remnants remains of interest. If the TeV flux of RX J1713.7-3946 is of neutral pion origin, then the accompanying charged pions will produce a guaranteed neutrino flux of roughly 10 muon-type neutrinos per kilometer-squared per year\cite{alvarezhalzen} and produce incontrovertible evidence for cosmic ray acceleration. Their calculation yields a result close to our previous estimate for a generic remnant. From a variety of such sources we can therefore expect event rates of cosmic neutrinos of galactic origin similar to those estimated for extragalactic neutrinos in the previous section. Supernovae associated with molecular clouds are a common feature of associations of OB stars that exist throughout the galactic plane, e.g. in the Cygnus region within view of IceCube.

It is important to realize that there is a robust relation between the neutrino and gamma flux emitted by cosmic ray accelerators\cite{alvarezhalzen}. It can also be exploited to estimate the neutrino flux from extragalactic sources. The $\nu_\mu + \bar\nu_\mu$ neutrino flux ($dN_\nu/dE_\nu$) produced by the decay of charged pions in the source can be derived from the observed gamma ray flux by energy conservation:
\begin{equation}
\int_{E_{\gamma}^{\rm min}}^{E_{\gamma}^{\rm max}}
E_\gamma {dN_\gamma\over dE_\gamma} dE_\gamma = K
\int_{E_{\nu}^{\rm min}}^{E_{\nu}^{\rm max}} E_\nu {dN_\nu\over dE_\nu} dE_\nu
\label{conservation}
\end{equation}
where ${E_{\gamma}^{\rm min}}$ ($E_{\gamma}^{\rm max}$) is the minimum (maximum) energy of the photons that have a hadronic origin. ${E_{\nu}^{\rm min}}$ and ${E_{\nu}^{\rm max}}$ are the corresponding minimum and maximum energy of the neutrinos. The factor $K$ depends on whether the $\pi^0$'s are of $pp$ or $p\gamma$ origin. Its value can be obtained from routine particle physics. In $pp$ interactions 1/3 of the proton energy goes into each pion flavor. In the pion-to-muon-to-electron decay chain 2 muon-neutrinos are produced with energy $E_\pi/4$ for every photon with energy $E_\pi/2$. Therefore the energy in neutrinos matches the energy in photons and $K=1$. The flux has to be reduced by a factor 2 because of oscillations. For $p\gamma$ interactions $K=1/4$. The estimate should be considered a lower limit because the observed photon flux to which the calculation is normalized may have been attenuated by absorption in the source or in the interstellar medium. 
 
In summary, the energetics of galactic as well as extragalactic cosmic rays points at the necessity to build kilometer-scale detectors to observe the associated neutrino fluxes that will reveal the sources. The case for doing neutrino astronomy with kilometer-scale instruments can also be made in other ways\cite{PR} and, as is usually the case, the estimates of the neutrino fluxes pointing at the necessity of such detectors are likely to be optimistic.

\section*{Acknowledgments}
I thank my IceCube collaborators as well as Luis Anchordoqui, Julia Becker, Concha Gonzalez-Garcia, Tom Gaisser, Aongus O'Murchadha and Grant Teply for discussions. This research was supported in part by the National Science Foundation under Grant No.~OPP-0236449, in part by the U.S.~Department of Energy under Grant No.~DE-FG02-95ER40896, and in part by the University of Wisconsin Research Committee with funds granted by the Wisconsin Alumni Research Foundation.

\end{document}